\def\papertitle{Aliasing Reduction in Neural Amp Modeling by Smoothing Activations}
\def\paperauthorA{Ryota Sato}
\def\paperauthorB{Julius O. Smith III}

\documentclass[twoside,a4paper]{article}
\usepackage{comment}
\usepackage{url}
\usepackage{etoolbox}
\usepackage{dafx_25}
\usepackage{amsmath,amssymb,amsfonts,amsthm}
\usepackage{euscript}
\usepackage[T1]{fontenc}
\usepackage[utf8]{inputenc}
\usepackage{ifpdf}
\usepackage[english]{babel}
\usepackage{caption}
\usepackage{subfig} 
\usepackage{color}

\input glyphtounicode
\ninept

\newcounter{numauth}
\newcounter{listcnt}
\newcommand\authcnt[1]{\ifdefined#1 \stepcounter{numauth} \fi}

\newcommand\addauth[1]{
\ifdefined#1 
\stepcounter{listcnt}
\ifnum \value{listcnt}<\value{numauth}
\appto\authorslist{, #1}
\else
\appto\authorslist{~and~#1}
\fi
\fi}
\authcnt{\paperauthorB}
\authcnt{\paperauthorC}
\authcnt{\paperauthorD}
\authcnt{\paperauthorE}
\authcnt{\paperauthorF}
\authcnt{\paperauthorG}
\authcnt{\paperauthorH}
\authcnt{\paperauthorI}
\authcnt{\paperauthorJ}
\def\authorslist{\paperauthorA}
\addauth{\paperauthorB}
\addauth{\paperauthorC}
\addauth{\paperauthorD}
\addauth{\paperauthorE}
\addauth{\paperauthorF}
\addauth{\paperauthorG}
\addauth{\paperauthorH}
\addauth{\paperauthorI}
\addauth{\paperauthorJ}

\usepackage{times}

\newif\ifpdf
\ifx\pdfoutput\relax
\else
   \ifcase\pdfoutput
      \pdffalse
   \else
      \pdftrue
   \fi
\fi

\ifpdf 
  \usepackage[pdftex,
    pdftitle={\papertitle},
    pdfauthor={\authorslist},
    pdfsubject={Proceedings of the 28th International Conference on Digital Audio Effects (DAFx25)},
    colorlinks=false, 
    bookmarksnumbered, 
    pdfstartview=XYZ 
  ]{hyperref}
  \usepackage[pdftex]{graphicx}
\else 
  \usepackage[dvips]{epsfig,graphicx}
  \usepackage[dvips,
    pdftitle={\papertitle},
    pdfauthor={\authorslist},
    pdfsubject={Proceedings of the 28th International Conference on Digital Audio Effects (DAFx25)},
    colorlinks=false, 
    bookmarksnumbered, 
    pdfstartview=XYZ 
  ]{hyperref}
\fi
\usepackage[hypcap=true]{caption}
\title{\papertitle}

\twoaffiliations{
\paperauthorA \,\thanks{\vspace{-3mm}}}
{\href{https://ee.stanford.edu/}{Dept. of Electrical Engineering} \\ Stanford University \\ Stanford, CA\\
{\tt \href{mailto:ryos17@stanford.edu}{ryos17@stanford.edu}}
}
{\paperauthorB \,\thanks{\vspace{-3mm}}}
{\href{https://dafx24.surrey.ac.uk}{Center for Computer Research in Music and Acoustics (CCRMA)} \\ Stanford University \\ Stanford, CA\\
{\tt \href{mailto:jos@ccrma.stanford.edu}{jos@ccrma.stanford.edu}}
}

\begin{document}
\ifpdf 
  \DeclareGraphicsExtensions{.png,.jpg,.pdf}
\else  
  \DeclareGraphicsExtensions{.eps}
\fi


\maketitle

\begin{abstract}
The increasing demand for high-quality digital emulations of analog audio hardware, such as vintage tube guitar amplifiers, led to numerous works on neural network-based black-box modeling, with deep learning architectures like WaveNet showing promising results. However, a key limitation in all of these models was the aliasing artifacts stemming from nonlinear activation functions in neural networks. In this paper, we investigated novel and modified activation functions aimed at mitigating aliasing within neural amplifier models. Supporting this, we introduced a novel metric, the Aliasing-to-Signal Ratio (ASR), which quantitatively assesses the level of aliasing with high accuracy. Measuring also the conventional Error-to-Signal Ratio (ESR), we conducted studies on a range of preexisting and modern activation functions with varying stretch factors. Our findings confirmed that activation functions with smoother curves tend to achieve lower ASR values, indicating a noticeable reduction in aliasing. Notably, this improvement in aliasing reduction was achievable without a substantial increase in ESR, demonstrating the potential for high modeling accuracy with reduced aliasing in neural amp models.
\end{abstract}

\section{Introduction}
\label{sec:intro}
Over the past few decades, \emph{virtual analog modeling} of audio
circuits has become a very active area of research \cite{virtualanalogeffects}, particularly in guitar amplifiers and effects. Digital clones of analog
amplifiers and effects pedals enable affordable mass production leading to revolutionary products like solid-state combo
amplifiers and multi-effects pedals \cite{dataintelo2024}.

\emph{White-box} models of analog audio circuits explicitly simulate all electrical components and their interconnections
\cite{wavesim,dunkel2016fender,massi2024wdf}. While effective,
these methods require deep circuit knowledge and attention
to detail.  Traditional \emph{black-box} methods such as Volterra series
and Wiener filters pose challenging system identification tasks
\cite{schattschneider1999discrete,PakarinenAndYeh09}.  More recently,
neural networks have proven effective for learning the input-output
map in black-box modeling \cite{martinez2020deep,moliner2025}.

\emph{Neural amp modeling}, particularly for nonlinear guitar tube
amplifiers, is a prominent field within black-box modeling.  Early efforts in neural amp
modeling utilized various recurrent neural networks (RNNs) such as
long-short-term memory (LSTM) RNN \cite{narx, lstm}, but the more
recent state-of-the-art methods are based on Temporal Convolutional
(Neural) Networks (TCNs) such as WaveNet \cite{wavenet}. These
models have demonstrated strong modeling capabilities at relatively
low computational cost, with very low Error-to-Signal Ratios
(ESR) \cite{Damskagg2018, alecwrightlstm}.  Relative to prior Wave
Digital Filter (WDF) amp models \cite{dunkel2016fender}, TCN amp
models suffer from significant aliasing artifacts, especially for high
fundamental frequencies \cite{alecwrightalias,
Damskgg2019RealTimeMO}. In TCNs, this aliasing is caused by the
neural \emph{activation functions} used, which are nonlinear and
foundational for neural networks.  It remains an open
challenge to eliminate audible aliasing at affordable computational
expense.

This study investigates the impact of \emph{activation function
choice} in mitigating aliasing artifacts in neural amplifier
models. To measure results, we introduce the Aliasing-to-Signal Ratio
(ASR), a novel metric designed based on number theory to most
accurately quantify aliasing in these models. We evaluate a range of
existing and custom activation functions to identify those that most
effectively reduce aliasing while maintaining model accuracy. Our
findings quantify the extent to which smoother activation functions
(e.g., due to larger ``stretch factors'') correspond to lower aliasing
levels, and at what increase in ESR. We find that certain activation
function families, such as Tanh and Snake, achieve the lowest
combinations of ASR and ESR.

The remainder of this paper is structured as follows:
Section \ref{sec:setup} describes the setup, including model
architecture, loss functions, training data, and model
configuration. Section \ref{sec:eval} presents the evaluation
methodology, focusing on ESR and the newly derived ASR
metrics. Section \ref{sec:experiments} details the experiments,
including the procedures to compare preexisting and newly constructed
activation functions, their tested results, a closer examination of
the Tanh and Snake functions with different stretch factors, and
analysis of its respective waveforms/spectra.

\section{Setup}
\label{sec:setup}

For our setup, we replicated the training environment present
in \cite{alecwrightalias}. For more detailed explanation of the setup, refer to that paper. 

\subsection{Model Architecture}

The model we utilized for training is the variant of
WaveNet \cite{wavenet} typically used in the neural amp modeling
community \cite{Damskagg2018}. Original WaveNet was an autoregressive
stack of dilated causal convolution layers (a 1D CNN inspired by
PixelCNN~\cite{oord2016conditionalimagegenerationpixelcnn}), with all
layer outputs feeding a classification head producing 8-bit
$\mu$-law-encoded samples (256 classification states). Two notable changes
introduced for neural amp modeling included: (1) a parallel,
non-recursive, feedforward WaveNet
variant~\cite{rethage2018wavenetspeechdenoising}, and (2) replacing
the classification head with a single linear 1×1 convolution (no
activations) applied to the sum of all layer
outputs~\cite{alecwrightalias}. It can also be called a Temporal
Convolutional Network (TCN).

\begin{figure}[ht]
\centerline{\includegraphics[scale=0.2]{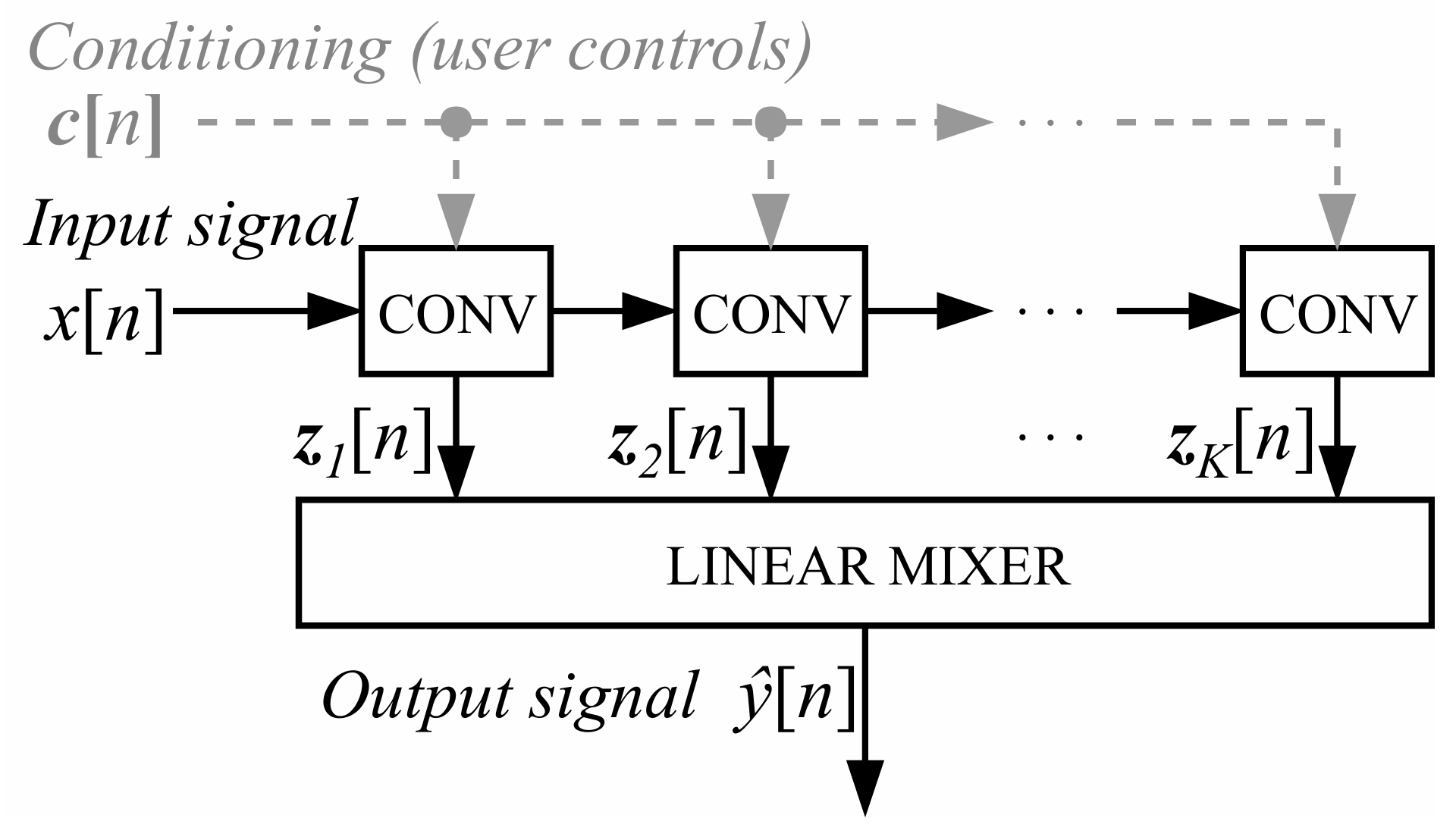}}
\caption{\label{wavenet architecture}{\it Current neural amp modeling architecture \cite{alecwrightalias}}}
\end{figure}

Figure \ref{wavenet architecture} shows the neural amp modeling
architecture from \cite{alecwrightalias}, which remains in wide use. The
input waveform samples $x[n]$ are fed to a series of dilated
convolution layers with channel dimension of $C$.  The outputs from each layer (\texttt{CONV} box) are
computed as learned FIR filters feeding nonlinear activation
functions.  The outputs from one layer to the next are called
``residual outputs,'' with channel dimension $C$, while the outputs
$z_k[n]$ are called ``skip outputs'' with the number of ``skip
channels'' being $C/2$ when the activations are gated and
$C$ otherwise.  Since the WaveNet classification head is replaced by a
linear down-projection to the output $\hat{y}[n]$, the activation
functions in the dilated convolutions \texttt{CONV} remain as the only
nonlinearities in the network.  In this paper,
we are interested in modifying these activation functions to reduce
aliasing. Optionally, as indicated by $c[n]$ in Figure ~\ref{wavenet
architecture}, we can condition the network with user controls such as
amplifier knob settings, but this paper does not explore that.

\subsection{Loss Function}
The model parameters were trained by minimizing the ``error-to-signal
ratio'' (ESR) with respect to the training data defined by
\[
\mathcal{E}_{\text{ESR}} = \frac{\displaystyle\sum_{n=0}^{N-1} |y_p[n] - \hat{y}_p[n]|^2}{\displaystyle\sum_{n=0}^{N-1} |y_p[n]|^2},
\]
where $y_p[n]$ is the pre-emphasized target signal and $\hat{y}_p[n]$
is the pre-emphasized model output. The pre-emphasis first-order high-pass filter, typically used in speech processing~\cite{Damskagg2018} and recent neural amp modeling works, is given by
\[
H(z) = 1 - 0.95z^{-1}.
\]

\subsection{Training Data}

For training data, we utilized the sample data provided by Steve
Atkinson's \href{https://www.neuralampmodeler.com/}{Neural Amp Modeler
(NAM)}, which provides
an \href{https://drive.google.com/file/d/1KbaS4oXXNEuh2aCPLwKrPdf5KFOjda8G/view}{input}-\href{https://drive.google.com/file/d/1NrpQLBbCDHyu0RPsne4YcjIpi5-rEP6w/view}{output}
pair for a heavy-distortion boutique tube amplifier. The input audio
file contains a variety of sounds that were chosen to maximize training
effectiveness (3 minutes 10 seconds in length), and it is widely used
in DIY neural amp modeling.

\subsection{Model Configuration}

For replication purposes, we followed the exact model configurations
present in \cite{alecwrightalias} other than the activation
functions. These parameters include a channel dimension of 16, kernel
size of 3, non-biased linear mixer (1x1 convolution), and an 18-layer
dilation pattern. This pattern, where $d_k$ represents the dilation rate for the $k$-th layer, is defined as follows
\[
d_k = \{1, 2, 4, \ldots, 256, 1, \ldots, 256\}.
\]

\section{Evaluation Method}
\label{sec:eval}

To evaluate the performances of our model, we utilized two metrics:
Error-to-Signal Ratio (ESR) as proposed in \cite{Damskagg2018} and
our new Aliasing-to-Signal Ratio (ASR).

\subsection{Error-to-Signal Ratio (ESR)}
The Error-to-Signal Ratio for evaluation is given by:

$$\mathcal{E}_{\text{ESR}} = \frac{\displaystyle\sum_{n=0}^{N-1} |y[n] - \hat{y}[n]|^2}{\displaystyle\sum_{n=0}^{N-1} |y[n]|^2}=\frac{P_{\text{error}}}{P_{\text{signal}}}$$

where $P_{\text{error}}$ is the power of the error signal (difference
between the output signal $\hat{y}[n]$ and target signal $y[n]$ with
$N$ number of samples) and $P_{\text{signal}}$ is the power of the
target signal.

\subsection{Aliasing-to-Signal Ratio (ASR)}
\label{sec:asr}

The Aliasing to Signal Ratio (ASR) provides an apparently novel
measure of the proportion of estimated aliasing energy $E_A$ in a real
signal $y[n]$ compared to the total harmonic energy $E_H$:
\[
\mathcal{E}_{\text{ASR}} = \frac{E_A}{E_H}
\]
where $E_A$ is defined as an estimate of the \emph{total aliased
energy}, while $E_H$ denotes the \emph{total harmonic energy}. More
specifically,
\begin{eqnarray*}
  E_H &=& \sum_{m=1}^{N_0} |Y(mk_0)|^2 \quad\mbox{(total harmonic energy)} \\
  E_A &=& E_Y - E_H \quad\mbox{(estimated total aliased energy)} \\
  E_Y &=& \sum_{k=0}^{(N-1)/2} |Y(k)|^2 \quad\mbox{(total spectrum energy)}
\end{eqnarray*}
where
\begin{itemize}
  \item $Y(k)$ is the discrete Fourier transform (DFT) of $y[n]$ at
    frequency bin $k = 0, 1, 2, \ldots, (N-1)/2$, using no windowing or zero padding
  \item $N$ is the DFT length, \emph{chosen to be relatively prime to $k_0$}
  \item $N_0=\lfloor\frac{N-1}{2k_0}\rfloor$ is the number of harmonic bins falling in the range $[1,(N-1)/2]$
  \item $k_0$ is the \emph{integer} DFT bin number corresponding to
    the test fundamental frequency $f_0 = f_s k_0/N$ in Hz
  \item $f_s$ is the sampling rate of $y[n]$ in Hz
\end{itemize}
The ASR quantifies the proportion of energy in the signal that comes
from aliasing artifacts. The closer the ASR is to zero, the less
aliasing is present in the signal.  It is defined as a linear ratio
(as opposed to dB) in order to compare more readily to the commonly
used ESR.

Because we use no windowing or zero-padding of the signal $y[n]$ prior
to the DFT, every DFT bin samples only a single frequency, and so can
be regarded as a set of discrete Fourier \emph{series} samples. This
representation remains valid for the nonlinearly processed test sine
provided that any transient response is discarded and the processed
signal is also periodic with the same period.

To allow $k_0$ to be any integer, we chose a large prime number for the DFT length $N$. We could alternatively let $N$ be any power of $2$, and restrict $k_0$ to be an odd integer. Either way, the sine-test fundamental frequency
$f_0=f_sk_0/N$, gives the property that all $N$ DFT bins receive
either a harmonic or aliasing component before any of the bins receive
a second component added in, starting at the $N$th harmonic, which is
a very high frequency where aliasing is typically negligible.  At this
point (the $N$th harmonic) the whole sequence of bin-filling repeats,
adding aliasing components first to the original harmonic bins
followed by adding to the rest of the bins.  This happens because the
set $\{k_0^n\}_{n=0}^{N-1}$ forms a \emph{complete residue system}
modulo $N$ when $k_0$ and $N$ are coprime.


The most audible aliasing occurs at high fundamental frequencies
$f_0$, so a representative test spectrum is very sparse, leaving the great
majority of bins for catching aliasing components created by nonlinear
amplifier models.  When the test fundamental $f_0$, sampling rate
$f_s$, and DFT size $N$ are chosen to be large, then a very good ASR
estimate is obtained.

In our tests, we chose $N=48,017$ corresponding to exactly one second
at $f_s=48,017$ Hz (the next prime after 48,000).  Our sine-test
duration was also set to one second to make the bin numbers
conveniently readable in Hz. In this case, there were no ``bin
collisions'' until after the $N$th harmonic at frequency $Nf_0 =
48,017 \cdot 1249 \approx 60$ MHz.  Since harmonic amplitudes roll off
fairly rapidly with frequency, thanks to the use of smooth activations
such as Tanh in NAM TCNs, the aliasing from that high up is presumed
negligible.


\section{Experiments}
\label{sec:experiments}
For running the experiments, we utilized an NVIDIA A100 GPU to speed
up the training process. Each model took approximately 1-2 minutes to
train, and was efficiently parallelized by training multiple models
with multiple GPUs at once.

\subsection{Activation Functions}
\label{sec:activation_function}

For our alias reduction experiments, we decided to test our models
with various activation functions. We employed all activation
functions present
in \href{https://pytorch.org/docs/stable/nn.html}{PyTorch's activation
function documentation}, and additional activation functions defined
below.

$$\text{Snake}(x) = x + \frac{1}{\alpha} \sin^2(\alpha x)$$

The Snake activation function is defined as above where $\alpha$ is a
positive parameter that controls the frequency of the sine wave
component. Higher values of $\alpha$ create more frequent oscillations
in the activation function. The Snake function maintains a consistent
derivative of 1 at $x = 0$ regardless of the value of $\alpha$. This
property helps maintain consistent gradient flow during training while
introducing nonlinearities that can capture complex patterns as seen
in \cite{kumar2023highfidelityaudiocompressionimproved}.

$$\text{ReLUSquared}(x) = \alpha \cdot (\max(0, x))^2$$

The ReLUSquared activation function applies a squaring operation to
the standard ReLU function, where $\alpha$ is a scaling
parameter. This function smoothes the nonlinear corner of ReLU,
thereby increasing the roll-off rate of aliasing components, and
quadratically emphasizes larger positive values. It excelled in LLM
sparsity tasks as seen in \cite{zhang2024relu2winsdiscoveringefficient}.

$$\text{ReLUSquaredDip}(x) = 
\begin{cases} 
x^2 & \text{if } x \geq 0 \\
\alpha \cdot x \cdot \sigma(x) & \text{if } x < 0
\end{cases}$$

The ReLUSquared with dip function combines properties of Swish and
ReLUSquared to create a function that transitions between cases,
where $\alpha$ is a scaling parameter for the Swish-like behavior in
the negative domain.

$$\text{Swish}(x) = x \cdot \sigma(x)$$

The Swish activation function, also known as SiLU (Sigmoid Linear
Unit), is defined as above where $\sigma(x)$ is the sigmoid function:
$\sigma(x) = 1/(1 + e^{-x})$. Swish is a smooth, non-monotonic
function that resembles ReLU but with a slight dip for negative
values. This non-monotonicity can help neural networks learn more
complex patterns compared to monotonic functions like ReLU as seen
in \cite{ramachandran2017searchingactivationfunctions}.

$$\text{Gaussian}(x) = e^{-x^2}$$

The Gaussian activation function applies a Gaussian transformation to the input.

$$\text{CustomTanh}(x) = \tanh\left(\frac{x}{\alpha}\right)$$

The CustomTanh activation function modifies the standard hyperbolic
tangent with a "stretch factor," where $\alpha$ is a positive parameter
that controls the horizontal stretching of the Tanh function. Larger
values of $\alpha$ make the function smoother and still preserves the
key properties of Tanh, including output range $(-1, 1)$ and
zero-centered activation.


\subsubsection{Gated Counterparts}
For each activation function, we also implemented a gated version
based on the gating mechanism originally used in
WaveNet \cite{wavenet} and also common in NAM.  This gated activation
form is defined as:

$$z = \text{Activation}(H_a x) \odot \sigma(H_g x)$$

where $\odot$ is the element-wise multiplication operation,
$\sigma(\cdot)$ is the sigmoid function, and $H_a$ and $H_g$ are
separate linear projections leading to the activation and gate,
respectively. Our implementation follows this structure, applying
different nonlinear activation functions in place of the
$\text{Activation}(\cdot)$ component while maintaining the sigmoid
gating mechanism.

\begin{figure*}[ht]
\center
\includegraphics[width=6in]{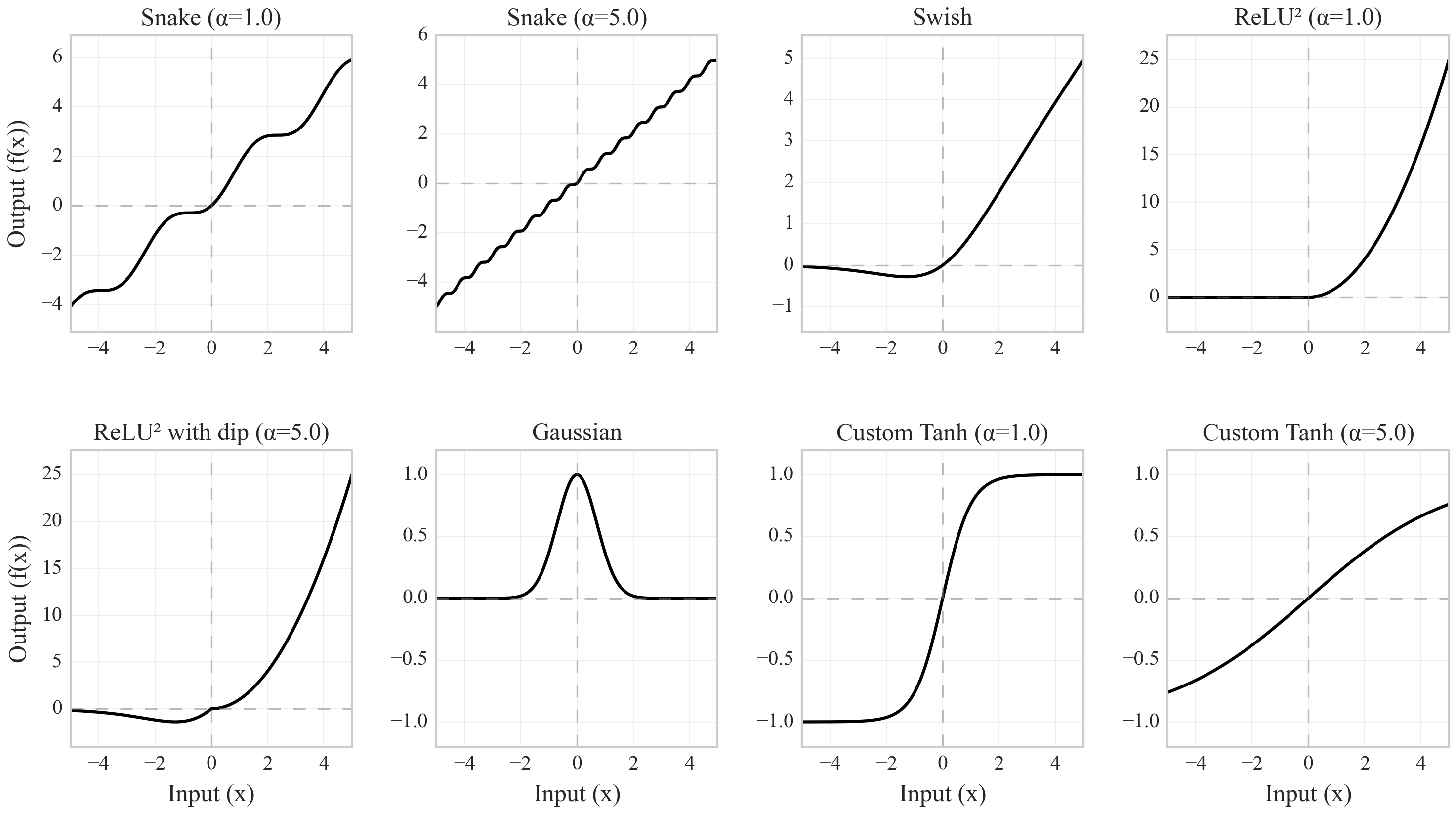}
\caption{\label{ftt_plot2}{\it Example activation functions mentioned in Section \ref{sec:activation_function}.}}
\end{figure*}

\subsection{General Results across All Activation Functions}
\label{sec:general_experiment}
To accurately test the performance of each activation function, we
conducted experiments using 100 deterministic training seeds for each
function and computed the average performance across all seeds. Each training run was capped at a maximum of 10 epochs for efficiency, with early stopping inherited from PyTorch Lightning to prevent overfitting. After
excluding activation functions from the default PyTorch library that
were incompatible with our framework, we evaluated a total of 62
unique activation functions. This set includes standard PyTorch
activation functions as well as parameterized versions of CustomTanh, ReluSquared, ReluSquaredDip,
and Snake, each tested with $\alpha$ scaling factors of 0.1, 0.2, 0.5,
1, 2, 4, 8, 16, and 32. In total, we examined 124 distinct
configurations, as each activation function was tested in both
non-gated and gated variants.

In our experimental nomenclature, activation functions are labeled
using a consistent format that conveys their configuration. Each
function is denoted by a prefix indicating whether it employs gating
(\texttt{True\_} for gated variants, \texttt{False\_} for non-gated),
followed by the activation function name
(e.g., \texttt{CustomTanh}, \texttt{Snake}, \texttt{ReluSquared}), and
finally a numeric value representing the scaling factor $\alpha$ where
applicable.

\subsubsection{Results}

\renewcommand{\arraystretch}{1.1}
\begin{table*}[ht]
  \caption{\itshape Comparison of Top Performing Models from Section \ref{sec:general_experiment}}
  \centering
  \footnotesize
  \setlength{\tabcolsep}{3pt}
  \begin{tabular}{|c|c|c|c|c||c|c|c|c|c|}\hline
    \multicolumn{5}{|c||}{\textbf{Top 10 by Average ASR}} & \multicolumn{5}{c|}{\textbf{Top 10 by Average ESR}} \\\hline
    \textbf{Activation Function} & \textbf{ASR} & \textbf{ASR std} & \textbf{ESR} & \textbf{ESR std} &
    \textbf{Activation Function} & \textbf{ASR} & \textbf{ASR std} & \textbf{ESR} & \textbf{ESR std} \\\hline
    \texttt{False\_CustomTanh\_32} & \textbf{0.001284} & 0.000554 & 0.089633 & 0.005450 &
    \texttt{True\_SELU} & 0.009103 & 0.010335 & \textbf{0.010591} & \textbf{0.001107} \\
    \texttt{False\_CustomTanh\_16} & 0.001319 & \textbf{0.000481} & 0.069236 & 0.005842 &
    \texttt{True\_CustomTanh\_0.5} & 0.005980 & 0.007825 & 0.011392 & 0.001202 \\
    \texttt{False\_CustomTanh\_8} & 0.001413 & 0.000660 & 0.044236 & 0.006788 &
    \texttt{True\_Hardtanh} & \textbf{0.004225} & 0.004222 & 0.011698 & 0.001134 \\
    \texttt{False\_CustomTanh\_2} & 0.001652 & 0.001604 & 0.017396 & 0.001916 &
    \texttt{True\_ELU} & 0.015245 & 0.018320 & 0.011699 & 0.001204 \\
    \texttt{False\_Snake\_4} & 0.001739 & 0.000519 & 0.016416 & 0.002005 &
    \texttt{True\_CELU} & 0.015245 & 0.018320 & 0.011699 & 0.001204 \\
    \texttt{False\_Snake\_8} & 0.001926 & 0.000599 & 0.020944 & 0.003570 &
    \texttt{True\_Snake\_2} & 0.015133 & 0.018640 & 0.012375 & 0.001375 \\
    \texttt{False\_CustomTanh\_1} & 0.002173 & 0.002096 & 0.013467 & 0.001799 &
    \texttt{False\_Hardtanh} & 0.002246 & \textbf{0.001579} & 0.012443 & 0.001285 \\
    \texttt{False\_Hardtanh} & 0.002246 & 0.001579 & \textbf{0.012443} & \textbf{0.001285} &
    \texttt{True\_CustomTanh\_1} & 0.005376 & 0.006392 & 0.012537 & 0.001326 \\
    \texttt{False\_Snake\_2} & 0.002320 & 0.001308 & 0.014682 & 0.001755 &
    \texttt{True\_Mish} & 0.012003 & 0.016054 & 0.012781 & 0.001410 \\
    \texttt{False\_CustomTanh\_4} & 0.002352 & 0.001922 & 0.025655 & 0.003470 &
    \texttt{True\_Snake\_1} & 0.016600 & 0.025875 & 0.012856 & 0.001562 \\\hline\hline
    \multicolumn{5}{|c||}{\textbf{Top 5 by Minimum ASR}} & \multicolumn{5}{c|}{\textbf{Top 5 by Minimum ESR}} \\\hline
    \textbf{Activation Function} & \textbf{ASR min} & \textbf{ASR std} & \textbf{ESR min} & \textbf{ESR std} &
    \textbf{Activation Function} & \textbf{ASR min} & \textbf{ASR std} & \textbf{ESR min} & \textbf{ESR std} \\\hline
    \texttt{False\_Sigmoid} & \textbf{0.000464} & 0.004420 & 0.026767 & 0.005238 &
    \texttt{True\_SELU} & \textbf{0.000817} & \textbf{0.010335} & \textbf{0.008176} & \textbf{0.001107} \\
    \texttt{False\_CustomTanh\_2} & 0.000520 & 0.001604 & 0.017145 & 0.001916 &
    \texttt{True\_CELU} & 0.054169 & 0.018320 & 0.008843 & 0.001204 \\
    \texttt{False\_Softsign} & 0.000544 & 0.003389 & \textbf{0.015599} & \textbf{0.001498} &
    \texttt{True\_ELU} & 0.054169 & 0.018320 & 0.008843 & 0.001204 \\
    \texttt{False\_CustomTanh\_32} & 0.000544 & \textbf{0.000554} & 0.091059 & 0.005450 &
    \texttt{True\_Snake\_2} & 0.057670 & 0.018640 & 0.009257 & 0.001375 \\
    \texttt{False\_CustomTanh\_4} & 0.000566 & 0.001922 & 0.026520 & 0.003470 &
    \texttt{True\_PReLU} & 0.002821 & 0.016104 & 0.009294 & 0.001561 \\\hline
  \end{tabular}
  \label{tab:combined_results}
\end{table*}

Upon training and evaluating on 12,400 models (124 activations each
with 100 seeds), we see interesting results shown in
Table \ref{tab:combined_results}. 

Looking at the average ASR performance, we observe that the top 10
activation functions were all non-gated variants. Specifically,
CustomTanh with larger stretch factors ($\alpha = 32, 16, 8$) achieves
remarkably low ASR values down to 0.001284, while maintaining ESR values below 0.10 (10\%
error). This matches our intuition that smoother, closer-to-linear
activation functions create less aliasing. Similarly, the Snake
activation function shows improved ASR performance with increased
oscillation frequency ($\alpha = 4, 8$), where the function approaches
a more linear, though modulated, behavior.
For CustomTanh, we see a decrease in ASR standard deviation as the
stretch factor increases (from $0.002096$ at $\alpha=1$ to $0.000554$
at $\alpha=32$). This suggests that CustomTanh with a larger stretch
factor produces good ASR models more consistently, which is another
benefit of stretching. Snake functions show a similar trend but with
generally lower standard deviations overall. For ESR standard
deviations, we see no visible pattern.

Conversely, when examining average ESR performance, we find that highly nonlinear
activation functions perform better at waveform
matching. Notably, gated variants dominate the top positions, with
\texttt{True\_SELU} achieving the lowest ESR of 0.010591. Compressed CustomTanh
(\texttt{True\_CustomTanh\_0.5}) performs particularly well with an ESR of
0.011392, demonstrating that more aggressively nonlinear activations can
effectively reduce error. Interestingly, ESR standard deviations closely follow their corresponding ESR values (lower ESR correlating with lower standard deviation), while ASR standard deviations show no such correlation with their ASR values. Additionally, examining the ASR metrics reveals
that these models introduce significantly more aliasing, with ASR
values approximately 6-9 times higher than the best-performing
configurations in terms of aliasing reduction. This illustrates the
interesting trade-off between ASR and ESR optimization, as the
characteristics that benefit one metric often come at the expense of
the other. The results suggest that while smoother, closer-to-linear
functions reduce aliasing, stronger nonlinearities are
better for waveform matching, model size being equal.

\subsubsection{Scatter Plot}

\begin{figure*}[ht]
\center
\includegraphics[width=7in]{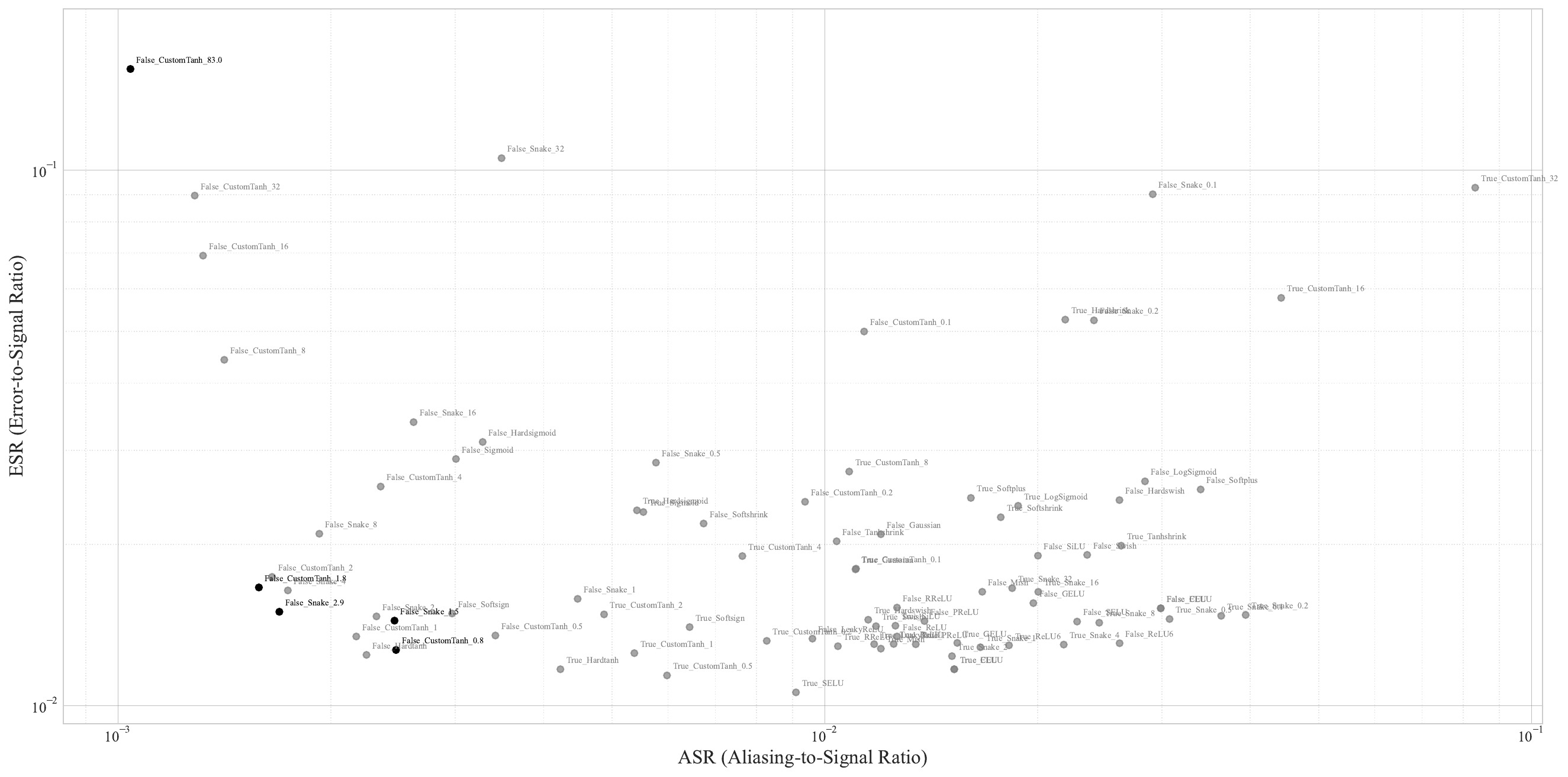}
\caption{\label{general_scatter}{\it Scatter plot of average ASR and ESR performance, excluding points for which ESR $>0.2$. Bold points represent notable models from Section \ref{sec:closer_experinent}}}
\end{figure*}

Figure \ref{general_scatter} provides a comprehensive visualization of
the ASR-ESR trade-off across all activation functions. Examining the
points closest to each axis reveals an inverse relationship between
ASR and ESR performance.  Reasonable non-gated activations to consider
for practical deployment include, progressing from less aliasing to
more, and more wave-matching error to less:\\
\texttt{False\_CustomTanh\_2}, \texttt{False\_Snake\_4}, \\
\texttt{False\_CustomTanh\_1} (commonly used now in
practice), and \texttt{False\_HardTanh}.

Discarding \texttt{False\_Snake\_4} until its aliasing spectral distribution
can be investigated, we are largely left with an elegant ungated Tanh
family (CustomTanh).  In this family, $\alpha=1$ serves as
the current standard default for neural amp modeling, while larger
$\alpha$ values such as $\alpha=2$ provide reduced aliasing at the
cost of increased wave-matching error. Conversely, smaller $\alpha$
values like $\alpha=1/2$ exhibit more aliasing but achieve more
precise waveform matching. Given side information about the lowest
pitch present in the input signal, the network could adaptively employ
high-alpha Tanh for high fundamentals (such as guitar solos high up
the neck), default Tanh for intermediate fundamentals, and
low-alpha Tanh for low fundamentals. Such control could be implemented
in real-time through a pedal or smoothed lower-bandlimit-follower.

Interesting extreme cases are observed near the upper left and right
of Figure \ref{general_scatter} with
\texttt{[False/True]\_CustomTanh\_32}. The stretch factor 32 makes the
activation function nearly linear, significantly reducing aliasing,
while increasing ESR to almost 10\%. The gated version demonstrates
that gating introduces high aliasing without improving modeling
accuracy at all in this case.

A nice surprise in Figure \ref{general_scatter}
is \texttt{False\_Hardtanh}, which is close to the Pareto optimal
boundary near the commonly used
\texttt{False\_CustomTanh\_1.0}, and showing a lower ESR with only
slightly more aliasing.  ``Hardtanh'' in PyTorch is a piecewise-linear
approximation to the Tanh function consisting of only three line
segments (flat, slope 1, and flat; or we could say ``zero-centered
clipped ReLU'').  We believe the Hardtanh family should be explored
using various slopes and smoothed corners of various curvatures, such
as can be obtained using cubic or higher-order polynomial splines.  To
avoid flat segments creating ``dead neurons,'' a small positive slope
can be added to the first and third segments, as in the PReLU
activation (ReLU with a slightly positively sloped cutoff segment).
Rounded corners on Hardtanh should reduce ASR while hopefully
preserving its superior ESR.

\subsubsection{Best Case Scenario (Minimum ASR and Minimum ESR)}

Since we tried 100 different seeds for generating random initial
weights, it is interesting to observe how much improvement can
be gained by taking the best
result. The bottom half of Table \ref{tab:combined_results} lists the best performing models
sorted by minimum ASR and ESR. The results reinforce our
earlier observations about the trade-off between ASR and ESR: the best
ASR performers are predominantly smooth, non-gated functions like
Sigmoid and CustomTanh, while the best ESR performers are exclusively
gated variants with more aggressive nonlinearities like SELU and
CELU. Note that for minimum ASR analysis, we excluded models such as \texttt{ReLUSquared} and \texttt{ReLUSquaredDip} that failed to train effectively (ESR $\approx$ 1). While these models achieved very low ASR values, this was likely due to near-silent output rather than meaningful aliasing reduction.

Intriguingly, the standard deviations reveal distinct patterns
between ASR and ESR metrics. For minimum ASR performance, there is no
discernible correlation between the minimum values and their
corresponding standard deviations (both for ASR and ESR), highlighting
the inherent difficulty in reliably optimizing for ASR. In contrast,
minimum ESR values show a clear correlation with their standard
deviations, where lower ESR values consistently correspond to lower
standard deviations. This pattern suggests that ESR optimization
exhibits more predictable behavior, while ASR performance appears to
be more sensitive to random initialization and requires careful
analysis in future research.

It is important to note that these minimum values represent the best
outcomes from random initialization rather than consistently
achievable performance. For example, while \texttt{True\_SELU} achieves an
impressive minimum ESR of 0.008176, its corresponding average ESR of
0.010591 is notably higher than the minimum case. Given that these
results are largely influenced by fortunate random initialization, we
consider the average performance metrics to be more reliable indicators of practical utility.

\subsection{Closer Examination on Selected Activation Functions}
\label{sec:closer_experinent}
To identify the activation function with minimal aliasing, we
focused on two function categories that achieved the lowest
average ASR values as shown in
Table \ref{tab:combined_results}: \texttt{False\_CustomTanh}
(non-gated Tanh with varying $\alpha$ values) and \texttt{False\_Snake}
(non-gated Snake with varying $\alpha$ values). For each activation
function, we evaluated 100 models with alpha values log-spaced between
$10^{-2}$ and $10^{2}$. To ensure statistical significance, each model
configuration was tested with 100 unique deterministic seeds.


Our comprehensive evaluation involved 20,000 models (2 activation
functions × 100 $\alpha$ values × 100 seeds), revealing distinct
patterns for each activation function. Notable models are listed in
Table \ref{tab:best_vs_baseline} and also plotted in bold in
Figure \ref{general_scatter} for comparison.

\begin{figure}[ht]
\centerline{\includegraphics[scale=0.55]{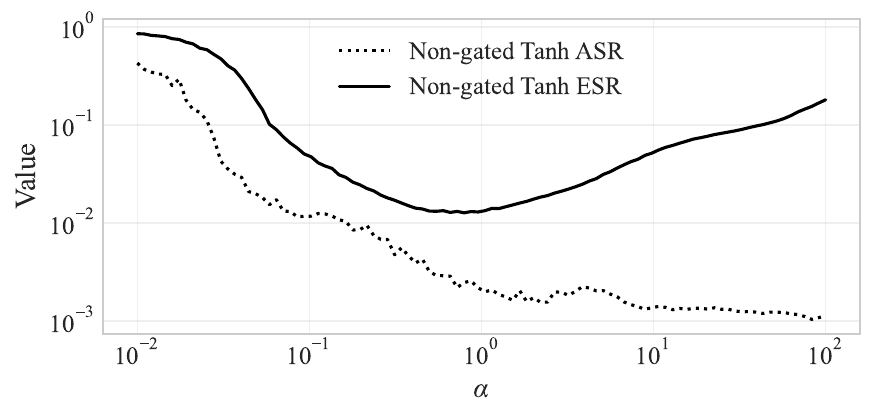}}
\caption{\label{tanh_comparison}{\it Detailed graph of $\alpha$ value vs ESR/ASR for Tanh.}}
\end{figure}

The Tanh activation function as seen in Figure \ref{tanh_comparison}
demonstrates that ESR follows a smooth convex curve with a local
minimum at approximately $\alpha=0.8$, suggesting that optimal modeling
capacity is achieved through a slightly horizontally compressed
non-gated Tanh function. The ASR exhibits a progressive decay with
$\alpha$, reaching a minimum of 0.001041, effectively halving the
aliasing compared to the baseline Tanh function ($\alpha=1$) as shown
in Table \ref{tab:best_vs_baseline}. For later tests, we chose
$\alpha=1.8$ to represent increased aliasing reduction with good
modeling accuracy, and $\alpha=0.8$ for best accuracy irrespective of
aliasing.

\begin{figure}[ht]
\centerline{\includegraphics[scale=0.55]{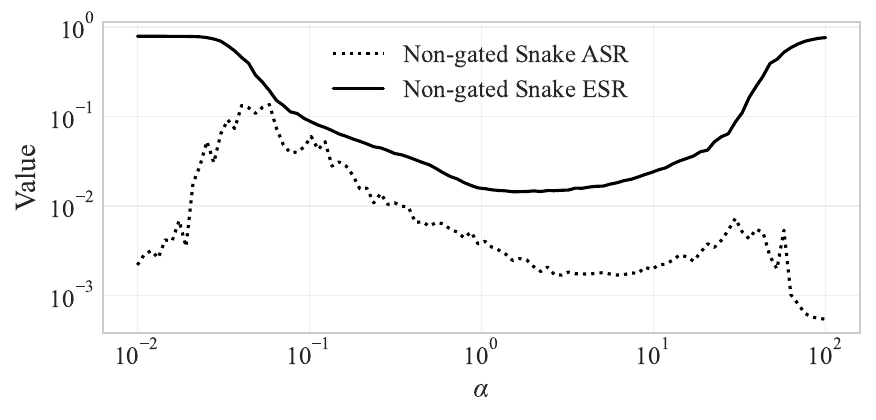}}
\caption{\label{snake_comparison}{\it Detailed graph of $\alpha$ value vs ESR/ASR for Snake.}}
\end{figure}

\begin{figure*}[ht]
\center
\includegraphics[width=7in]{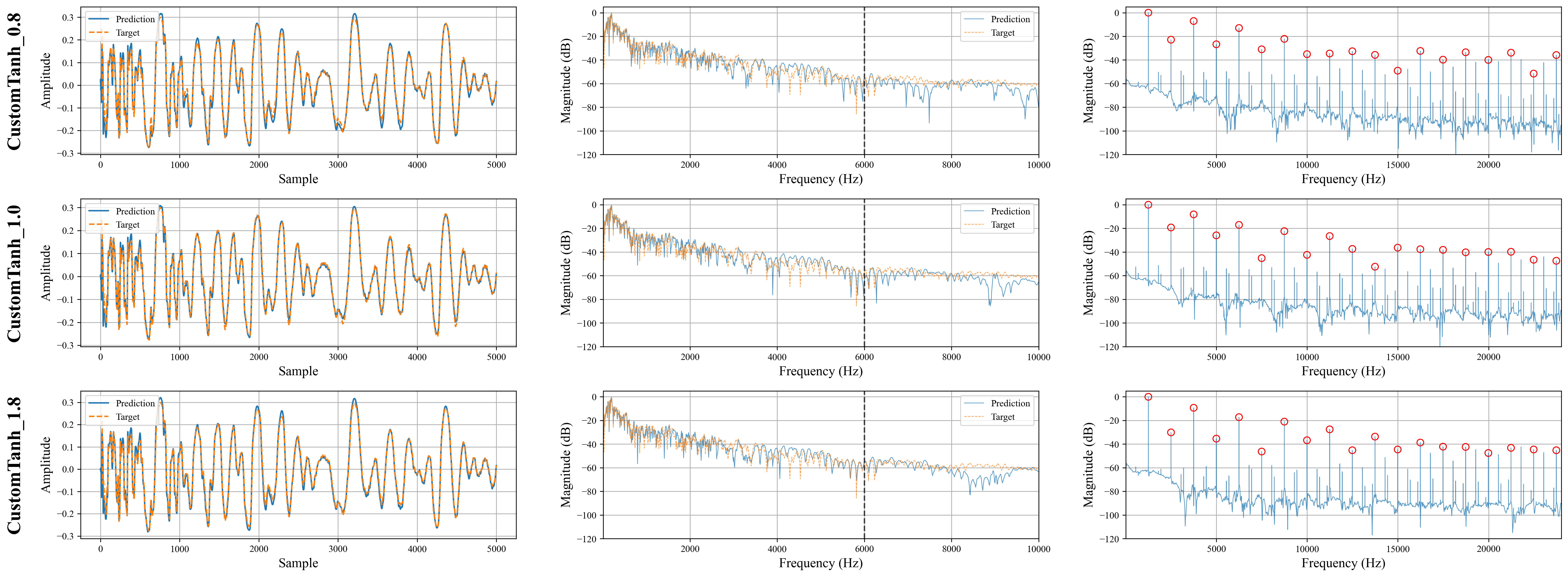}
\caption{\label{waveform_spectrum_analysis}{\it Comparing prediction
 and target output waveform (left column), prediction and target
 spectra, with a black dashed at 6kHz indicating a typical
 guitar-cabinet bandwidth (center column), and sine-wave spectrum with
 harmonics circled (right column).}}
\end{figure*}

The Snake activation function exhibits similar ESR behavior as seen in
Figure \ref{snake_comparison} with a convex region and broad global
minimum near $\alpha=1.5$ at only slightly less accuracy than
Tanh. Its ASR shows a similar convex zone over the $\alpha$ range
corresponding to ESR below 10\%, or roughly $\alpha\in[0.1,40]$.  The
very-low ASR regions at the extremes coincide with ESR values
approaching 1, indicating severely compromised modeling
capability---likely producing near-silent audio output. This suggests
that the optimal Snake configuration lies near $\alpha=2.9$, where the
local minimum for ASR occurs within the broad ESR minimum.

Our analysis reveals fundamental differences between these activation
functions. Snake's behavior shows that increasing $\alpha$
(oscillation frequency) affects aliasing with clear minima for both
ESR and ASR, indicating an optimal point balancing performance and
aliasing reduction. In contrast, Tanh exhibits progressive ASR decay,
demonstrating that smoother variants further reduce aliasing while
maintaining reasonable modeling capability. Although our experiments
were limited to $\alpha \leq 10^2$, we hypothesize that higher
$\alpha$ values would further reduce ASR while increasing ESR.  We
also hypothesize that, for a given ASR, the ESR can be improved by
increasing the channel dimension (from 16 to 32, 64, or 128) as
demonstrated in \cite{alecwrightalias}, trading more processing speed
for quality. Given the significant advances in computing power
since \cite{alecwrightalias} was published (which used an Apple iMac
with a 2.8GHz Intel Core i5 processor), it is now more affordable to
increase the model order (channel dimension, convolution kernel
length, and or number of convolution layers, etc.)  for better ESR
performance.

\begin{table}[ht]
  \caption{\itshape Comparison of Best Performing Models with Baselines.}
  \centering
  \footnotesize
  \begin{tabular}{|c|c|c|}
    \hline
    \textbf{Activation Function} & \textbf{ASR} & \textbf{ESR} \\\hline
    \texttt{False\_CustomTanh\_1} (\textit{Baseline}) & 0.002173 & 0.013467 \\
    \texttt{False\_CustomTanh\_83.0} &\textbf{ 0.001041} & 0.154460 \\
    \texttt{False\_CustomTanh\_0.8} & 0.002473 & \textbf{0.012708} \\\hline
    \texttt{False\_Snake\_1} (\textit{Baseline}) & 0.004469 & 0.015831 \\
    \texttt{False\_Snake\_100} & \textbf{0.000544} & 0.759755 \\
    \texttt{False\_Snake\_1.5} & 0.002461 & \textbf{0.014415} \\\hline
    \texttt{False\_CustomTanh\_1.8} (\textit{Balanced}) & 0.001582 & 0.016628 \\
    \texttt{False\_Snake\_2.9} (\textit{Balanced}) & 0.001691 & 0.014975 \\\hline
  \end{tabular}
  \label{tab:best_vs_baseline}
\end{table}

\subsection{Waveform and Spectrum Analysis}

To further validate our results, we conducted an in-depth
analysis of the CustomTanh function variants, focusing on their
modeling characteristics as presented in
Table~\ref{tab:best_vs_baseline}. We
excluded \newline \texttt{False\_CustomTanh\_83.0} due to
its relatively high ESR. The specific model instances were selected
based on their normalized distance from the mean, choosing the seed
that minimized the sum of normalized ASR and ESR distances as shown in
Table~\ref{tab:best_vs_baseline}. Figure~\ref{waveform_spectrum_analysis}
presents an overlay of the three CustomTanh variants $\alpha\in\{0.8,
1.0, 1.8\}$.
Each row corresponds to a different model configuration, with three
distinct visualizations per model.

The left column displays waveform comparisons between the ground truth
(orange dashed line) and the model prediction (blue solid line). Notably,
all three variants demonstrate comparable waveform modeling capacity,
with no large deviations in the audio waveform approximation.

The center column shows the magnitude-spectrum overlays for the ground truth and model prediction up to $10$ kHz, with a vertical dashed line at 6~kHz
indicating the typical upper bandlimit of guitar speaker cabinets. The
flat behavior of the target spectrum as $10$ kHz is approached
continues with a slight decline out to the Nyquist limit ($24$~kHz),
and the predicted spectrum stayed well below that. For all three models, we observe
a similar matching spectral contour below 6~kHz, indicating comparable
model quality. Above 6~kHz, the prediction spectra show varying
degrees of deviation below the target, a pattern that continues out to
$24$~kHz. Future research could explore several approaches to better handle high-frequency content: (1) applying post-processing low-pass filters to simulate cabinet response, (2) utilizing higher quality training data with stronger pre-emphasis (e.g., +12 dB/octave) to improve modeling above the cabinet corner frequency, and (3) developing perceptually-weighted loss functions that account for the ear's reduced frequency resolution at higher frequencies, penalizing only auditorily relevant spectral envelope deviations.

The right column, designed to highlight aliasing as
in \cite[Fig.~12]{Damskagg2019}, presents a sine-wave test using a
1249~Hz fundamental frequency sampled at 48,017~Hz, with harmonics
circled in red to distinguish them from aliasing components (uncircled
peaks).
Analysis of the aliasing components below $6$~kHz reveals a
systematic relationship between the CustomTanh stretch factor $\alpha$
and aliasing suppression. CustomTanh with $\alpha=0.8$ exhibits
consistent aliasing components at approximately $-50$~dB, while
$\alpha=1.0$ shows improvement down to $\approx$ $-55$~dB. The most
effective aliasing suppression is achieved by $\alpha=1.8$, where
aliasing remains below $-60$~dB.

\section{Conclusions}
This work investigates the relationship between smooth activation functions and aliasing in neural amp modeling, introducing the Aliasing-to-Signal Ratio (ASR) to quantify aliasing artifacts. 

We found that smoother, ungated activation functions consistently
produce less aliasing, with the non-gated CustomTanh family emerging
as a particularly flexible and effective choice. By adjusting the
stretch factor $\alpha$, CustomTanh offers a continuous trade-off
between aliasing reduction and modeling accuracy, i.e., increasing
$\alpha$ generally decreases ASR while increasing ESR. 
Notably, \texttt{CustomTanh\_1.8} achieves aliasing
components below $-60$~dB (approximately 27\% decrease in aliasing
compared to the baseline model), while maintaining acceptable ESR
performance.  


In contrast, less smooth activation functions, particularly gated
variants, excel at minimizing ESR (with \texttt{True\_SELU} achieving the
lowest ESR of 0.010591, approximately 21\% decrease in signal error
compared to the baseline model). However, they consistently introduce
more aliasing, following the inherent trade-off between ASR and
ESR. Future research should include listening tests to perceptually
validate these results and determine optimal operating
points for different applications.

Our work demonstrates that thoughtful selection of activation functions can significantly reduce aliasing in neural amp models without requiring architectural changes or additional computational overhead. Future research directions include:
\begin{itemize}
    \item Exploring higher channel dimensions to improve ESR while maintaining the anti-aliasing benefits of smoother activation functions
    \item Developing hybrid loss functions that explicitly minimize both ASR and ESR
          for sinusoidal training samples
          
    \item Investigating separately learnable activation parameters at the network, layer, or neuron levels
    \item Examining additional activation function families beyond Tanh and Snake to further optimize the aliasing-reduction versus modeling-accuracy trade-off
\end{itemize}

\section{Acknowledgment}
Thanks to the \href{https://research.reazon.jp/projects/ReazonSpeech/}{ReazonSpeech}
team from Reazon Holdings for providing computational resources!

\nocite{*}
\bibliographystyle{IEEEbib}
{ 
\bibliography{DAFx25_tmpl} 

\begin{thebibliography}{10}

\bibitem{virtualanalogeffects}
V.~Välimäki, S.~Bilbao, J.~O. Smith, J.~S. Abel, J.~Pakarinen, and D.~Berners,
\newblock {\em DAFx: Digital Audio Effects}, chapter 12: Virtual Analog Effects, pp. 473--522,
\newblock John Wiley \& Sons, Ltd, 2011.

\bibitem{dataintelo2024}
DataIntelo,
\newblock ``Multi effects pedals market report | global forecast from 2025 to 2033,'' 2024.

\bibitem{wavesim}
M.~Karjalainen and J.~Pakarinen,
\newblock ``Wave digital simulation of a vacuum-tube amplifier,''
\newblock in {\em 2006 IEEE International Conference on Acoustics Speech and Signal Processing Proceedings}, 2006, vol.~5, pp. V--V.

\bibitem{dunkel2016fender}
W.~R. Dunkel, M.~Rest, K.~J. Werner, M.~J. Olsen, and J.~O.~Smith III,
\newblock ``The {Fender Bassman 5F6-A} family of preamplifier circuits---a wave digital filter case study,''
\newblock in {\em DAFx-16}, Sept. 2016.

\bibitem{massi2024wdf}
O.~Massi, E.~Manino, and A.~Bernardini,
\newblock ``Wave digital modeling of circuits with multiple one-port nonlinearities based on {Lipschitz}-bounded neural networks,''
\newblock in {\em DAFx-24}, Sept. 2024.

\bibitem{schattschneider1999discrete}
J.~Schattschneider and U.~Z{\"o}lzer,
\newblock ``Discrete-time models for non-linear audio systems,''
\newblock in {\em DAFx-99}, Dec. 1999.

\bibitem{PakarinenAndYeh09}
J.~Pakarinen and D.~T. Yeh,
\newblock ``A review on digital guitar tube amplifier modeling techniques,''
\newblock {\em Computer Music Journal}, vol. 33, pp. 85--100, 2009,
\newblock {\footnotesize http://www.mitpressjournals.org/doi/pdf/10.1162/comj.2009.33.2.85}.

\bibitem{martinez2020deep}
M.~A.~Mart{\'\i}nez Ram{\'\i}rez, E.~Benetos, and J.~D. Reiss,
\newblock ``Deep learning for black-box modeling of audio effects,''
\newblock {\em Applied Sciences}, vol. 10, no. 2, pp. 638, 2020.

\bibitem{moliner2025}
E.~Moliner, M.~Švento, A.~Wright, L.~Juvela, P.~Rajmic, and V.~Välimäki,
\newblock ``Unsupervised estimation of nonlinear audio effects: Comparing diffusion-based and adversarial approaches,'' 2025,
\newblock arXiv:2504.04751 [eess.AS].

\bibitem{narx}
J.~Covert and D.~L. Livingston,
\newblock ``A vacuum-tube guitar amplifier model using a recurrent neural network,''
\newblock in {\em 2013 Proceedings of IEEE Southeastcon}, 2013, pp. 1--5.

\bibitem{lstm}
Z.~Zhang, E.~Olbrych, J.~Bruchalski, T.~J. McCormick, and D.~L. Livingston,
\newblock ``A vacuum-tube guitar amplifier model using long/short-term memory networks,''
\newblock in {\em SoutheastCon 2018}, 2018, pp. 1--5.

\bibitem{wavenet}
A.~van~den Oord, S.~Dieleman, H.~Zen, K.~Simonyan, O.~Vinyals, A.~Graves, N.~Kalchbrenner, A.~Senior, and K.~Kavukcuoglu,
\newblock ``{WaveNet}: A generative model for raw audio,'' 2016,
\newblock arXiv:1609.03499 [cs.SD].

\bibitem{Damskagg2018}
E.-P. Damsk{\"a}gg, L.~Juvela, E.~Thuillier, and V.~V{\"a}lim{\"a}ki,
\newblock ``Deep learning for tube amplifier emulation,'' 2018,
\newblock arXiv:1811.00334 [eess.AS].

\bibitem{alecwrightlstm}
A.~Wright, E.-P. Damskägg, and V.~Välimäki,
\newblock ``Real-time black-box modelling with recurrent neural networks,''
\newblock in {\em DAFx-19}, Sept. 2019.

\bibitem{alecwrightalias}
A.~Wright, E.-P. Damskägg, L.~Juvela, and V.~Välimäki,
\newblock ``Real-time guitar amplifier emulation with deep learning,''
\newblock {\em Applied Sciences}, vol. 10, no. 3, 2020.

\bibitem{Damskgg2019RealTimeMO}
E.-P. Damsk{\"a}gg, L.~Juvela, and V.~V{\"a}lim{\"a}ki,
\newblock ``Real-time modeling of audio distortion circuits with deep learning,''
\newblock in {\em DAFx-19}, Sept. 2019.

\bibitem{oord2016conditionalimagegenerationpixelcnn}
A.~van~den Oord, N.~Kalchbrenner, O.~Vinyals, L.~Espeholt, A.~Graves, and K.~Kavukcuoglu,
\newblock ``Conditional image generation with pixelcnn decoders,'' 2016,
\newblock arXiv:1606.05328 [cs.CV].

\bibitem{rethage2018wavenetspeechdenoising}
D.~Rethage, J.~Pons, and X.~Serra,
\newblock ``A wavenet for speech denoising,'' 2018,
\newblock arXiv:1706.07162 [cs.SD].

\bibitem{kumar2023highfidelityaudiocompressionimproved}
R.~Kumar, P.~Seetharaman, A.~Luebs, I.~Kumar, and K.~Kumar,
\newblock ``High-fidelity audio compression with improved {RVQGAN},'' 2023,
\newblock arXiv:2306.06546 [cs.SD].

\bibitem{zhang2024relu2winsdiscoveringefficient}
Z.~Zhang, Y.~Song, G.~Yu, X.~Han, Y.~Lin, C.~Xiao, C.~Song, Z.~Liu, Z.~Mi, and M.~Sun,
\newblock ``{ReLU}$^2$ wins: Discovering efficient activation functions for sparse {LLM}s,'' 2024,
\newblock arXiv:2402.03804 [cs.LG].

\bibitem{ramachandran2017searchingactivationfunctions}
P.~Ramachandran, B.~Zoph, and Q.~V. Le,
\newblock ``Searching for activation functions,'' 2017,
\newblock arXiv:1710.05941 [cs.NE].

\bibitem{Damskagg2019}
E.-P. Damskägg, L.~Juvela, and V.~Välimäki,
\newblock ``Real-time modeling of audio distortion circuits with deep learning,''
\newblock in {\em SMC 2019}, May 2019, pp. 332--339.

\bibitem{xu2025kernel}
Y.~Xu, B.~Mu, and T.~Chen,
\newblock ``On kernel design for regularized volterra series identification of {Wiener--Hammerstein} systems,''
\newblock {\em preprint}, May 2025,
\newblock arXiv:2505.20747 [eess.SY].

\end{thebibliography}
}

\end{document}